\documentclass[review]{elsarticle}

\usepackage{lineno,hyperref}
\modulolinenumbers[5]

\journal{Journal of \LaTeX\ Templates}

\biboptions{numbers,sort&compress}

\begin{document}

\begin{frontmatter}

\title{Quartet structure of $N=Z$ nuclei in a boson formalism: the case of $^{28}$Si}

\author[address1]{M. Sambataro\corref{correspondingauthor}}
\cortext[correspondingauthor]{Corresponding author}
\ead{michelangelo.sambataro@ct.infn.it}

\author[address2]{N. Sandulescu}
\ead{sandulescu@theory.nipne.ro}

\address[address1]{Istituto Nazionale di Fisica Nucleare - Sezione di Catania, Via S. Sofia 64, I-95123 Catania, Italy}
\address[address2]{National Institute of Physics and Nuclear Engineering, P.O. Box MG-6, Magurele, Bucharest, Romania}

\begin{abstract}
The structure of the $N=Z$ nucleus $^{28}$Si is studied by resorting to an IBM-type formalism with $s$ and $d$ bosons representing isospin $T=0$ and angular momentum $J=0$ and $J=2$ quartets, respectively. $T=0$ quartets are four-body correlated structures formed by two protons and two neutrons. The microscopic nature of the quartet bosons, meant as images of the fermionic quartets, is investigated by making use of a mapping procedure and is supported by the close resemblance between the phenomenological and microscopically derived Hamiltonians. 
The ground state band and two low-lying side bands, a $\beta$ and a $\gamma$ band, together with all known $E2$ transitions and quadrupole moments associated with these states are well reproduced by the model. 
An analysis of the potential energy surface 
places $^{28}$Si, only known case so far, at the critical point of the U(5)-$\overline{\rm SU(3)}$ transition of the IBM structural diagram.
\end{abstract}


\end{frontmatter}


\section{Introduction}

The important role played by quartets in $N=Z$ nuclei has been known 
for a long time \cite{flowers,ginocchio,faraggi,arima,yamamura,catara}. By quartets we
denote here alpha-like four-body correlated structures formed by two protons and two
neutrons coupled to total isospin $T=0$. Recently, microscopic quartet models have been successfully employed to describe the proton-neutron pairing \cite{chasman,qcm_t1,qm_t1a,qm_t1b,qcm_def_t0t1,qm_t0t1} as well as general two-body interactions 
\cite{qm_prla,qm_prlb,qm_odd,qcm_gen} in $N=Z$ nuclei.
As a basic outcome, the $J=0$ quartet has been found to play a leading role but  other low-$J$ quartets have also been found essential to describe the
spectra of $N=Z$ nuclei.

The difficulties associated with a microscopic treatment of $N=Z$ nuclei in a formalism of quartets rapidly grow with increasing the number of active nucleons. To make
the application of this formalism possible also for large systems, in the present work we propose an approach where elementary bosons replace quartets. Based upon the above
fermionic studies, we search for a description of $N=Z$ nuclei in terms of only two building blocks, the $T=0,J=0$  and $T=0,J=2$ quartets. These quartets are therefore represented as elementary $s$ and $d$ bosons, respectively. 
This bosonic architecture clearly coincides 
with that of the Interacting Boson Model (IBM) in its simplest 
version \cite{iachebook}. The application of this model, in terms of quartets, to $N=Z$ nuclei is, however, without precedent. We remark that in the standard IBM framework a proper treatment of even-even $N=Z$ nuclei would imply the use of the much more elaborate 
IBM-4 version of the model \cite{elliott}
which carries 10 different types of pair bosons.
We also notice that an IBM-type approach based on quartet bosons was applied long ago \cite{dukelsky}, on 
a purely phenomenological basis, to nuclei with protons and neutrons 
occupying different major shells, i.e. nuclei which are commonly 
described by IBM-2 \cite{iachebook}.
   
The manuscript is structured as follows. In Section 2, we illustrate our formalism.
In Section 3, this formalism is applied to a description of $^{28}$Si.
In Section 4, we discuss the geometric structure of this nucleus.
Finally, in Section 5, we give the  conclusions.

\section{The formalism}

We start by setting the general quartet boson formalism for the treatment of $N=Z$ nuclei. We describe these nuclei in terms of collective $T=0$ ($J=0$ and 2) quartets that we represent as elementary $sd$ bosons. 
By denoting the corresponding boson creation operators as
$b^\dagger_0=s^\dagger$ and $b^\dagger_{2\mu}=d^\dagger_\mu$ ($\mu$ being the angular momentum projection),  
the most general one-body plus two-body Hamiltonian takes the standard IBM form 
\begin{eqnarray}
&&H_B=\sum_\lambda \epsilon (\lambda)\hat{n}_\lambda +\nonumber\\
&&\sum_{\lambda_1\lambda_2,\lambda '_1\lambda '_2,\Lambda}
V({\lambda_1\lambda_2,\lambda '_1\lambda '_2;\Lambda})
[[b^\dagger_{\lambda_1}b^\dagger_{\lambda_2}]^{\Lambda}
[{\tilde b}_{\lambda '_1}{\tilde b}_{\lambda '_2}]^{\Lambda}]^{0}
\label{1}
\end{eqnarray}
where $\hat{n}_\lambda =\sum_\mu b^\dagger_{\lambda\mu}b_{\lambda\mu}$ and
$\tilde{b}_{\lambda\mu}=(-1)^{\lambda+\mu}b_{\lambda -\mu}$.
To evaluate to what extent the quartet bosons can be associated to microscopic quartets as well as to have an initial guess for the parameters of this Hamiltonian we shall resort to a mapping procedure.
Mapping procedures allow to establish a link between spaces of composite and elementary objects and have been largely employed in a microscopic analysis of the IBM \cite{klein}. 
In this work we will follow the general lines of the procedure of Ref. \cite{sambamap} adapted for the quartet case.

We begin by introducing the most general quartet with isospin $T=0$ and angular momentum (projection) $J(M)$  
\begin{eqnarray}
Q^\dagger_{JM}&=&\sum_{i_1j_1J_1}\sum_{i_2j_2J_2}\sum_{T'}
C^{(J)}_{i_1j_1J_1,i_2j_2J_2,T'}\nonumber\\
&&\times\Bigl[ [a^\dagger_{i_1}a^\dagger_{j_1}]^{J_1T'}[a^\dagger_{i_2}a^\dagger_{j_2}]^{J_2T'}\Bigr]^{J,T=0}_{M}.
\label{3}
\end{eqnarray}
With $N$ such quartets 
we construct the fermionic quartet space
\begin{equation}
F^{(N)}=\{Q^\dagger_{i_1} Q^\dagger_{i_2} \cdots Q^\dagger_{i_N} |0\rangle\}_
{i_1\leq i_2\dots \leq i_N},
\label{4}
\end{equation}
where $Q^\dagger_i\equiv Q^\dagger_{J_iM_i}$.
To the quartet operator $Q^\dagger_i$ we associate the boson $b^\dagger_i$ and,
in correspondence with the fermion space $F^{(N)}$, we define the boson space 
\begin{equation}
B^{(N)}=({{\cal N}_{i_1i_2\dots i_N}})^{-1/2}
b^\dagger_{i_1} b^\dagger_{i_2} \cdots b^\dagger_{i_N} |0)\}_
{i_1\leq i_2\dots \leq i_N}.
\label{5}
\end{equation}
where  ${\cal N}_{i_1i_2\dots i_N}$ is a normalization factor. There is a one-to-one correspondence between the
states of $F^{(N)}$ and $B^{(N)}$, the basic difference being that the boson states are orthonormal while the fermion ones are not. In correspondence with a fermion Hamiltonian $H_F$, we define a boson hamiltonian $H_B$ such that
\begin{equation}
(N,l|H_B|N,m)=\sum_{ij}R^{(N)}_{li}\langle N,i|H_F|N,j\rangle R^{(N)}_{jm}
\label{6}
\end{equation}
where $|N,i\rangle$ and $|N,i)$ are generic states of $F^{(N)}$ and $B^{(N)}$, respectively, and 
$R^{(N)}_{li}={\sum_k}^*f^{(N)}_{lk}{{\cal N}^{(N)}_k}^{-1/2}f^{(N)}_{ik}$
with $f^{(N)}_{lk}$ and ${\cal N}^{(N)}_k$ being the eigenfuntions and eigenvalues of the overlap matrix of the fermion states $|N,i\rangle$, respectively. The asterisk in the expression for $R^{(N)}_{li}$ means that the sum is extended only over the non-zero eigenvalues ${\cal N}^{(N)}_k$. It can be proved that the eigenspectrum of $H_B$ contains all the eigenvalues of $H_F$ in $F^{(N)}$ plus a number of zero's
corresponding to the states with ${\cal N}^{(N)}_k =0$ .
The boson Hamiltonian $H_B$ so constructed is Hermitian and, in general, $N$-body. 
Analogous expressions for $H_B$, but for pair bosons, can be found in Refs. \cite{skouras,johnson}.

\section{The spectrum of $^{28}$Si}

We apply the formalism just described to the nucleus $^{28}$Si. $^{28}$Si has 6 protons and 6 neutrons outside the $^{16}$O core. 
Thus we describe this nucleus in terms of three collective quartets 
that we represent as elementary $sd$ bosons. 
The corresponding theoretical spectrum has only 10 states. The angular momenta of the states are such that these can be arranged
into a ground state band and two side bands, a $\beta$ and a $\gamma$ band.
Correspondingly, as experimental spectrum of $^{28}$Si we consider only the ground state band and two low-lying $\beta$ and $\gamma$ bands. 
These $\beta$ and $\gamma$ bands have their band heads at 4.98 MeV and 7.42 MeV, respectively. 
According to Ref. \cite{sheline}, these ground, $\beta$ and $\gamma$ bands share a common intrinsic structure, all being classified as ``oblate".
These experimental bands are shown on the left side of Fig. 1.
Some uncertainties are present for the  $J=4$ state of the $\beta$ band due to the lack of experimental information.
The state which has been tentatively inserted in Fig. 1 is the $J=4$ state at $E=10.67$ MeV. It is worth mentioning that
the experimental spectrum shown in Fig. 1 is only a part of the complex spectrum of $^{28}$Si, which contains many other bands \cite{sheline}.
\begin{figure}[ht]
\begin{center}
\includegraphics[width=4.2in,height=4.9in,angle=-90]{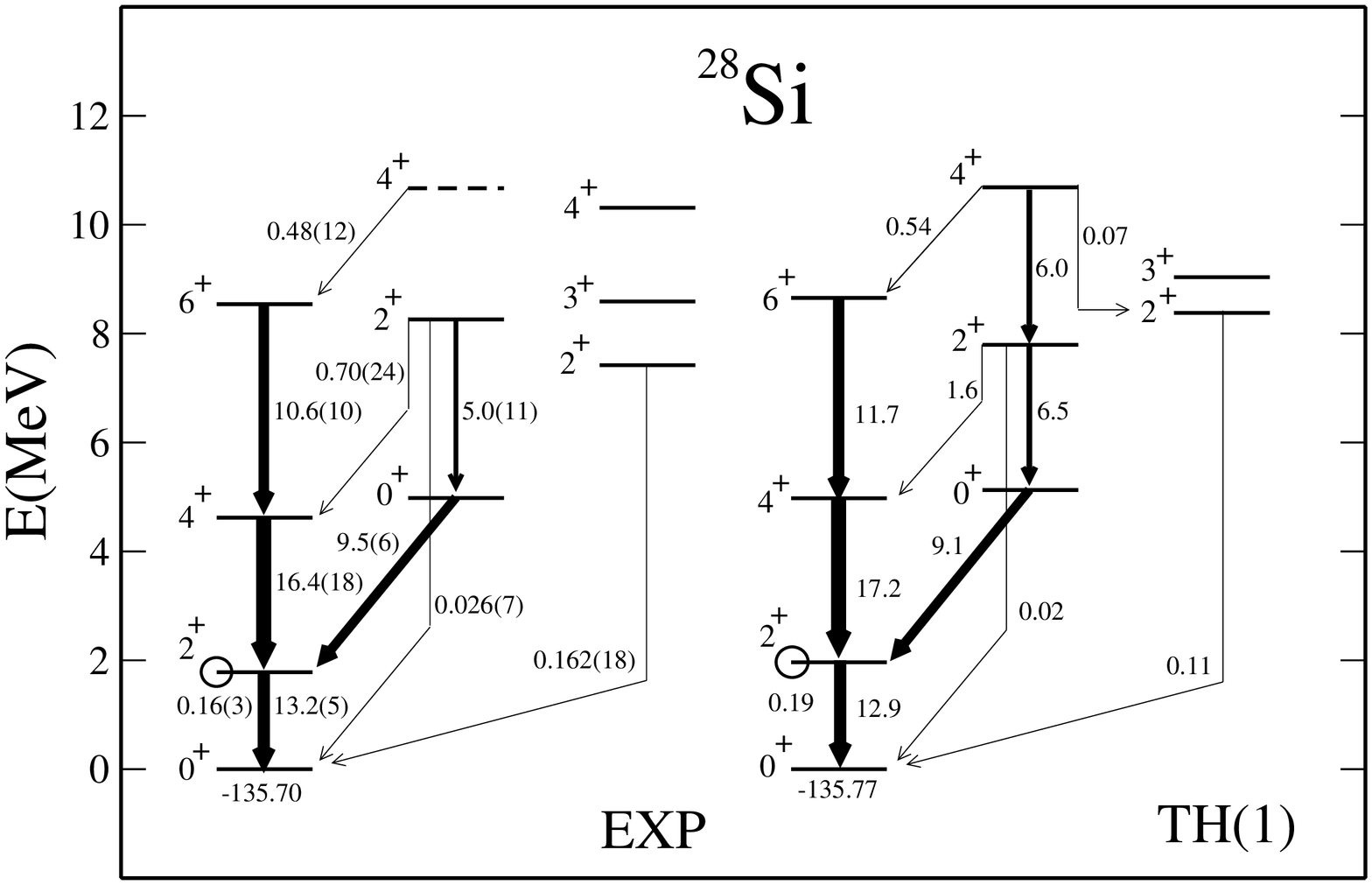}
\bf{Figure 1}
\end{center}
\end{figure}

To describe the spectrum of Fig. 1 with the Hamiltonian (1)
we proceed as follows. Consistently with a microscopic interpretation of the bosons, as energies of the $s$ and $d$ bosons 
we adopt the energies of the lowest $J=0$ and $J=2$ states resulting from a shell model calculation for a system of two protons and two neutrons in the $sd$ shell. The values obtained using the USDB interaction \cite{brown} are $\epsilon (0)=-37.713$ MeV and $\epsilon (2)=-36.158$ MeV. The remaining parameters, i.e. the two-body matrix elements of the Hamiltonian, are fitted to the experimental data. As a starting point for this fit we have used the two-body matrix elements derived from the USDB interaction according to the boson mapping 
presented above. These matrix elements are shown in Fig. 2  (dashed line).  In the same figure we show (solid line) the matrix elements which provide the best fit of the experimental spectrum. With the notation of Fig. 2, the adopted values are (in MeV): (1)=-3.374, (2)=-0.859, (3)=-3.875, (4)=-14.298, (5)=-2.348, (6)=-6.746, (7)=-9.316. 
Some differences can be seen between microscopically derived and phenomenologically fitted parameters (particularly at point (3)).
These differences, which have significant effects on the final spectrum, are expected to reflect a renormalization of the boson parameters which takes into account the lack of $J>2$ quartets (whose role has been previously pointed out \cite{qm_prla}) as well as the lack of three-body terms in $H_B$. The overall agreement between the two set of parameters of Fig. 2 is, however, such to support the microscopic interpretation of the bosons as images of $T=0$ quartets.
\begin{figure}[ht]
\begin{center}
\includegraphics[width=4.2in,height=4.9in,angle=-90]{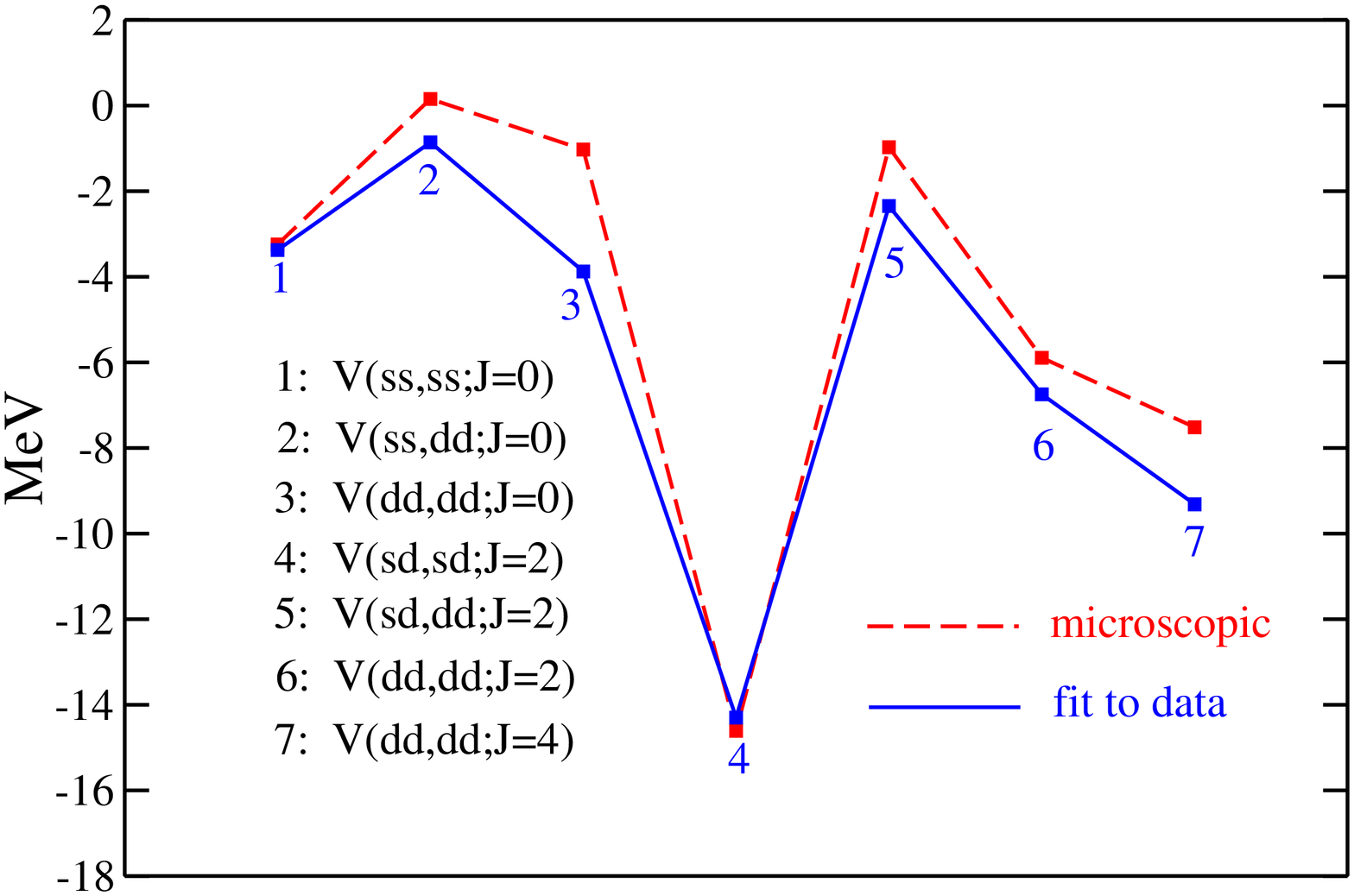}
\bf{Figure 2}
\end{center}
\end{figure}

The theoretical spectrum of the Hamiltonian (1) with the parameters fitted as discussed above is shown Fig. 1. A good agreement is seen  between theory and experiment.  
The calculations generate also a $0^+$ state at 11.61 MeV, not shown in the figure. The only certain experimental $0^+$ state in this region is located at 10.27 MeV \cite{exp} but
there are no sufficient elements to establish a clear connection between these two states.
In the same figure one can also notice that the theoretical second $J=4$ state belongs the $\beta$ band, as testified by the reported $B(E2)$'s. 

To evaluate the $E2$ transitions we have adopted the standard IBM operator
$T^{(E2)}_\mu=e_B( [d^\dagger_\mu s + s^\dagger \tilde{d}_\mu]^2+
\chi [d^\dagger\tilde{d}]^2_\mu )$
with  $e_B =1.45$ (W.u.)$^{1/2}$ and $\chi =2.1$. As seen in Fig.1, the agreement for the $B(E2)$ values is quite good. This agreement indicates that not only the energies but also the wave functions of the low-lying states shown in Fig. 1  are well
described in the present formalism of quartet bosons. This represents a fundamental prerequisite for the analysis of the geometry of $^{28}$Si that we are going to illustrate.

\section{The geometric structure of $^{28}$Si}

Any IBM-type Hamiltonian has associated with it an intrinsic geometric structure \cite{ginocchio2,dieperink}. This is defined by means of the coherent state
\begin{equation}
| N;\beta ,\gamma\rangle =({N!(1+\beta^2)^N})^{-1/2}(B^\dagger )^N|0\rangle
\label{8}
\end{equation}
where
$B^\dagger =s^\dagger +\beta [{\rm cos}\gamma~ d^\dagger _0+2^{-1/2}
{\rm sin}\gamma~ (d^\dagger _{+2}+d^\dagger _{-2})]$.
The variables $\beta$ and $\gamma$ identify the intrinsic shape of the nucleus. $\gamma =0^{\rm o}$ corresponds to a prolate deformation while 
$\gamma =60^{\rm o}$ to an oblate one.
The equilibrium shape of the nucleus is defined by the values $\beta_0$, $\gamma_0$ which minimize the potential energy surface
$E(\beta ,\gamma)=\langle N;\beta ,\gamma |H_B| N;\beta ,\gamma\rangle$.
\begin{figure}[ht]
\begin{center}
\includegraphics[width=4.2in,height=4.9in,angle=-90]{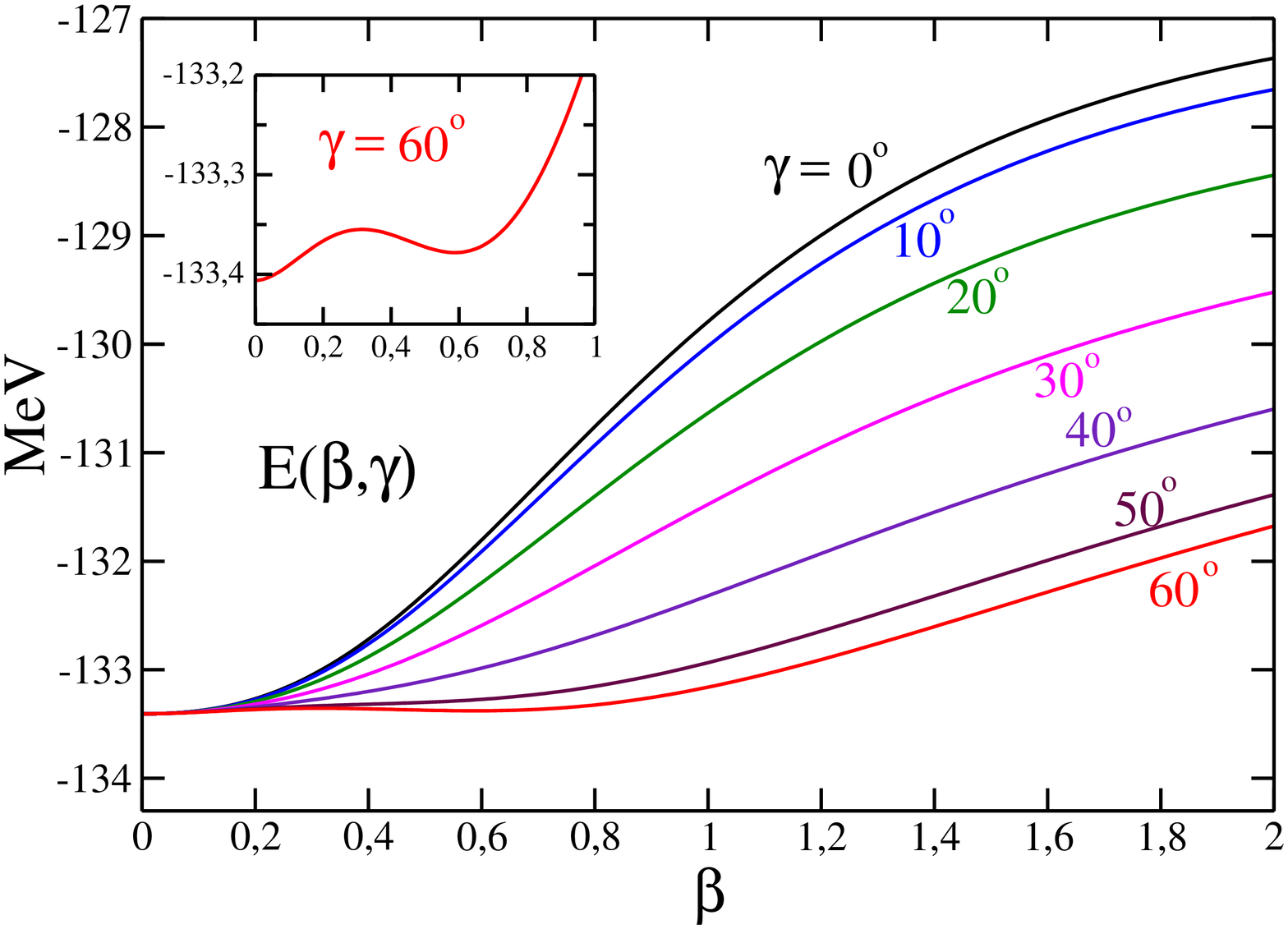}
\bf{Figure 3}
\end{center}
\end{figure}
In Fig. 3 we show $E(\beta ,\gamma)$ for $^{28}$Si, obtained with the Hamiltonian (\ref{1}), for different values of $\gamma$. At first sight $E(\beta ,\gamma )$ appears to be characterized by a very flat region  which extends over a broad range of $\beta$ ($0-0.8$) in the $\gamma =60^{\rm o}$ section. A more accurate inspection, however, reveals the true peculiarity of this behavior, namely the existence of two almost degenerate minima  at $\beta =0$ (the lowest one) and $\beta\simeq 0.6$ (see the inset of Fig. 3). The ground state of  $^{28}$Si therefore emerges from this analysis as characterized by a coexistence of spherical and oblate shapes.

Previous studies of the potential energy surface  have evidenced that the presence of two shallow coexisting minima tipically occurs at the critical point of a first order phase transition \cite{iache}.
To illustrate this point we introduce the schematic Hamiltonian \cite{iache,jolie}
\begin{equation}
H^{(T)}_B=(1-\eta )\widehat{n}_d-\eta (Q^\dagger\cdot Q^\dagger),
\label{11}
\end{equation}
where $Q^\dagger=[d^\dagger s+s^\dagger\tilde{d}]^{(2)}+\chi [d^\dagger\tilde{d}]^{(2)}$ and 
$\chi  =+\frac{\sqrt{7}}{2}$.
In this Hamiltonian the control parameter $\eta$ allows to move continuously 
from the spherical U(5) limit to the oblate $\overline{\rm SU(3)}$ limit.
In between these two limits the system undergoes a first order phase transition at $\eta _{cr}=0.129$. This criticality emerges from a discontinuity at $\eta _{cr}$ in the first derivative of the minimum of the potential energy surface $E(\beta ,\gamma)$ similarly to what found in the U(5)-SU(3) case \cite{iachebook,cejnar}.

\begin{figure}[ht]
\begin{center}
\includegraphics[width=4.2in,height=4.9in,angle=-90]{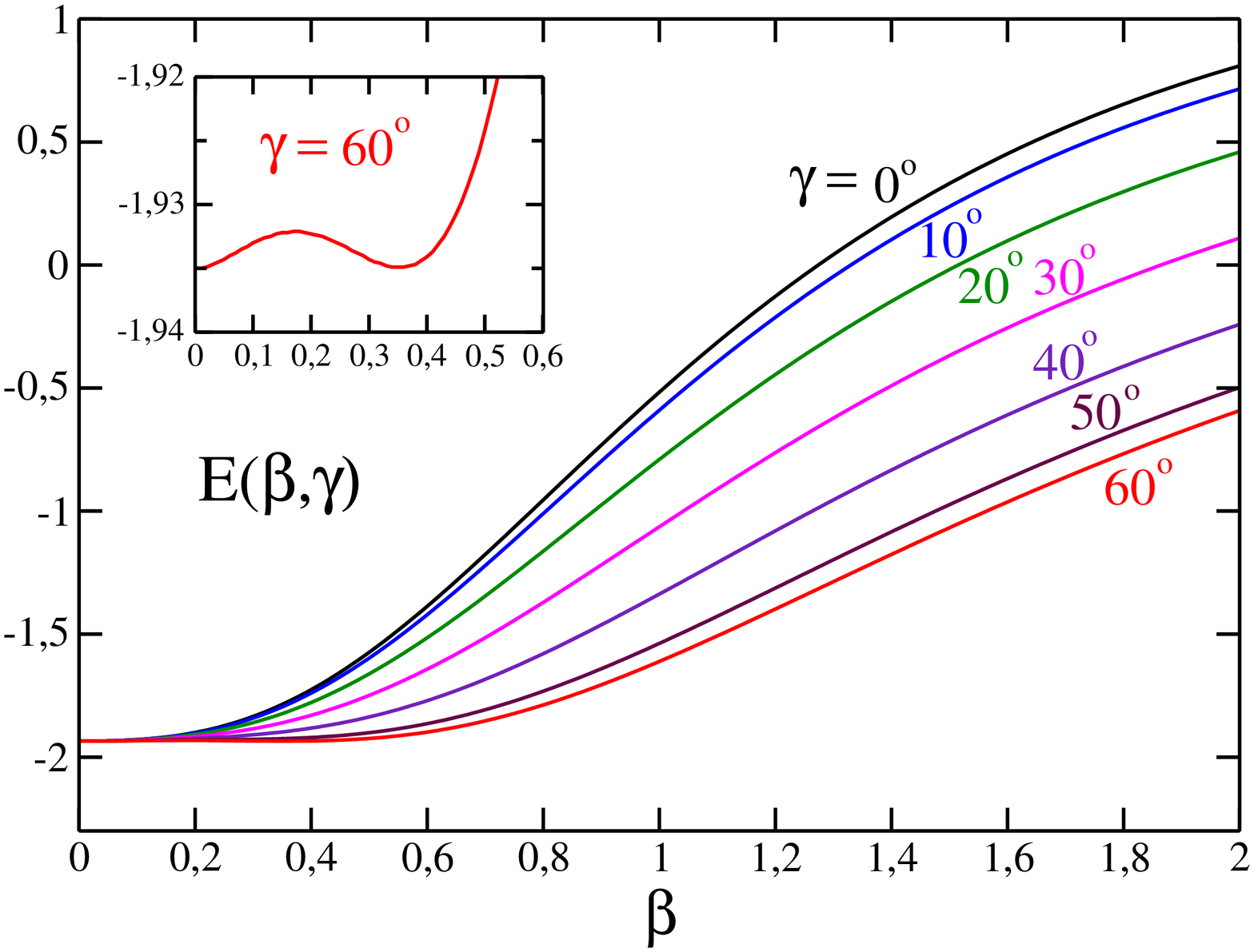}
\bf{Figure 4}
\end{center}
\end{figure}

In Fig. 4 we show the potential energy surface associated with $H^{(T)}_B$ at $\eta =\eta _{cr}$. One sees that the curve for $\gamma=60^{\rm o}$ 
shows a coexistence of a spherical and an oblate minimum quite similar to that 
of Fig. 3, the two minima being in this case exactly degenerate.
The two minima observed in 
Fig. 4 for $\gamma =60^{\rm o}$  coexist
(although in a non-degenerate form) in a very narrow interval around $\eta _{cr}$
(0.127-0.143). Outside this interval only one minimum is left either at $\beta =0$  (for $\eta <0.127$) or at $\beta >0.5$ (for $\eta >0.143$). On the basis of the above evidences one can clearly identify the potential energy surface of $^{28}$Si as that of a nucleus at the
critical point of the U(5)-$\overline{\rm SU(3)}$ transition . 

Stimulated by the similarity of the potential energies of $H_B$  and $H^{(T)}_B$, we have explored to which extent the schematic Hamiltonian (\ref{11}) is able to mimic the full Hamiltonian (\ref{1}).  We have therefore evaluated the spectrum of $H^{(T)}_B$ at $\eta =\eta _{cr}$. To make the comparison meaningful,  $H^{(T)}_B$ has been multiplied by the scale factor $\rho =2.45$ (MeV). We have also adopted the so-called consistent $Q$-formalism \cite{warner} therefore formulating the $E2$ operator in terms of the same $Q^\dagger$ defining the quadrupole-quadrupole interaction of $H^{(T)}_B$. This has been scaled by the factor $\tau =1.68$ (W.u.)$^{1/2}$.

In Fig. 5 we compare the results obtained with the Hamiltonians (\ref{1}) and (\ref{11}).
Consistently with the observed similarity between the potential energies, the schematic Hamiltonian is found to reproduce to a reasonable extent the results of the full Hamiltonian (\ref{1}) in spite
of the fact that no parameters have been involved in this comparison (except for scale factors). 
\begin{figure}[ht]
\begin{center}
\includegraphics[width=4.2in,height=4.9in,angle=-90]{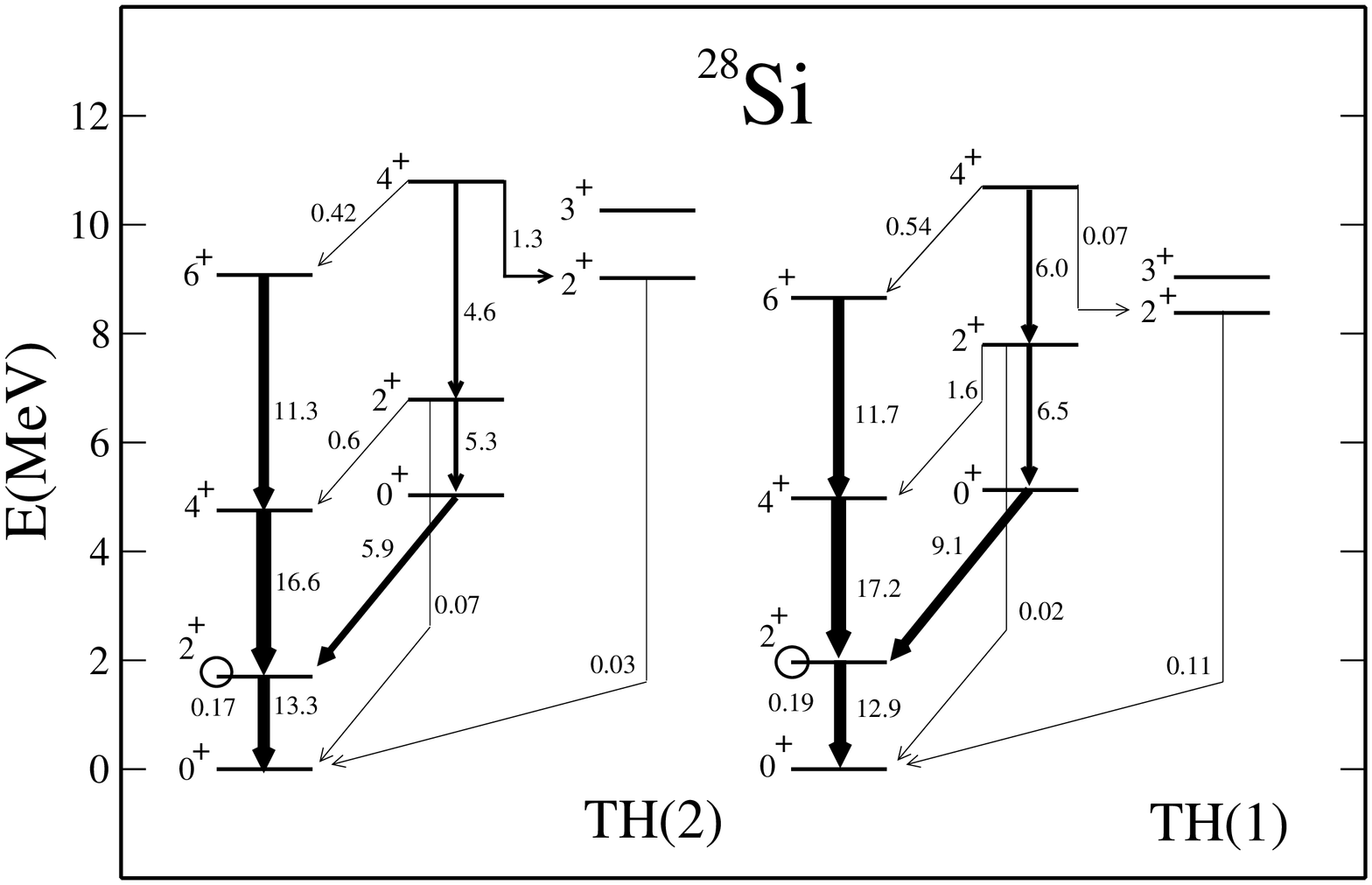}
\bf{Figure 5}
\end{center}
\end{figure}

In what follows we will explore whether precursors of a phase 
transition can be observed in a system of only $N=3$ bosons, such as
$^{28}$Si.
To do so it is appropriate to look at observables that
are particularly sensitive to the control parameter in the critical region.
In Fig. 6 we show two such observables. The first one, shown in the main panel, 
is the ratio $B(E2;0^+_2-2^+_1)/B(E2;2^+_1-0^+_1)$. The experimental value of
this ratio, 0.72(0.05), corresponds to a value of $\eta$ close to 
$\eta_{cr}$. From Fig. 6 one can notice that this B(E2) ratio strongly depends
on the control parameter $\eta$ and this dependence further grows with increasing 
$N$ (inset (a)). For $N=10$ one observes a precipitous drop in this ratio which marks 
the occurrence of a phase transition. As it can also be inferred from a glance at the inset (a),
$\eta_{cr}$ significantly decreases with increasing $N$. However, such a 
decrease can be very much smoothed out if the strength of the quadrupole-quadrupole term
in the Hamiltonian (\ref{11}) is rescaled by a factor $4N$, as often done in the
literature \cite{iache}.

In Fig. 6 we also show (inset (b)) the ratio 
$R_{42}=E(4^+_1)/E(2^+_1)$. For $^{28}$Si
the experimental value for this ratio is equal to 2.6 
and this too corresponds to a value $\eta$ close to $\eta_{cr}$.
As seen in Fig. 6, this energy ratio strongly depends on 
$\eta$ already for $N=3$. This dependence becomes more marked for larger 
$N$ and, for $N=10$, the abrupt increase of $R_{42}$ with $\eta$ signals a 
phase transition between the vibrational ($R_{42}=2$) and rotational 
($R_{42}=3.3$) limits.
In conclusion, the experimental values of both observables shown in Fig. 6 point to a 
value of $\eta$ compatible with $\eta_{cr}$. In addition, the transitional patterns
of these observables clearly exhibit the occurrence of a phase transition in systems with rather large $N$. For $N=3$, these patterns are somewhat smoothed out owing to the smallness of the system but they still show significant fingerprints of a  precursor of
a phase transition in $^{28}$Si. 
\begin{figure}[ht]
\begin{center}
\includegraphics[width=4.2in,height=4.9in,angle=-90]{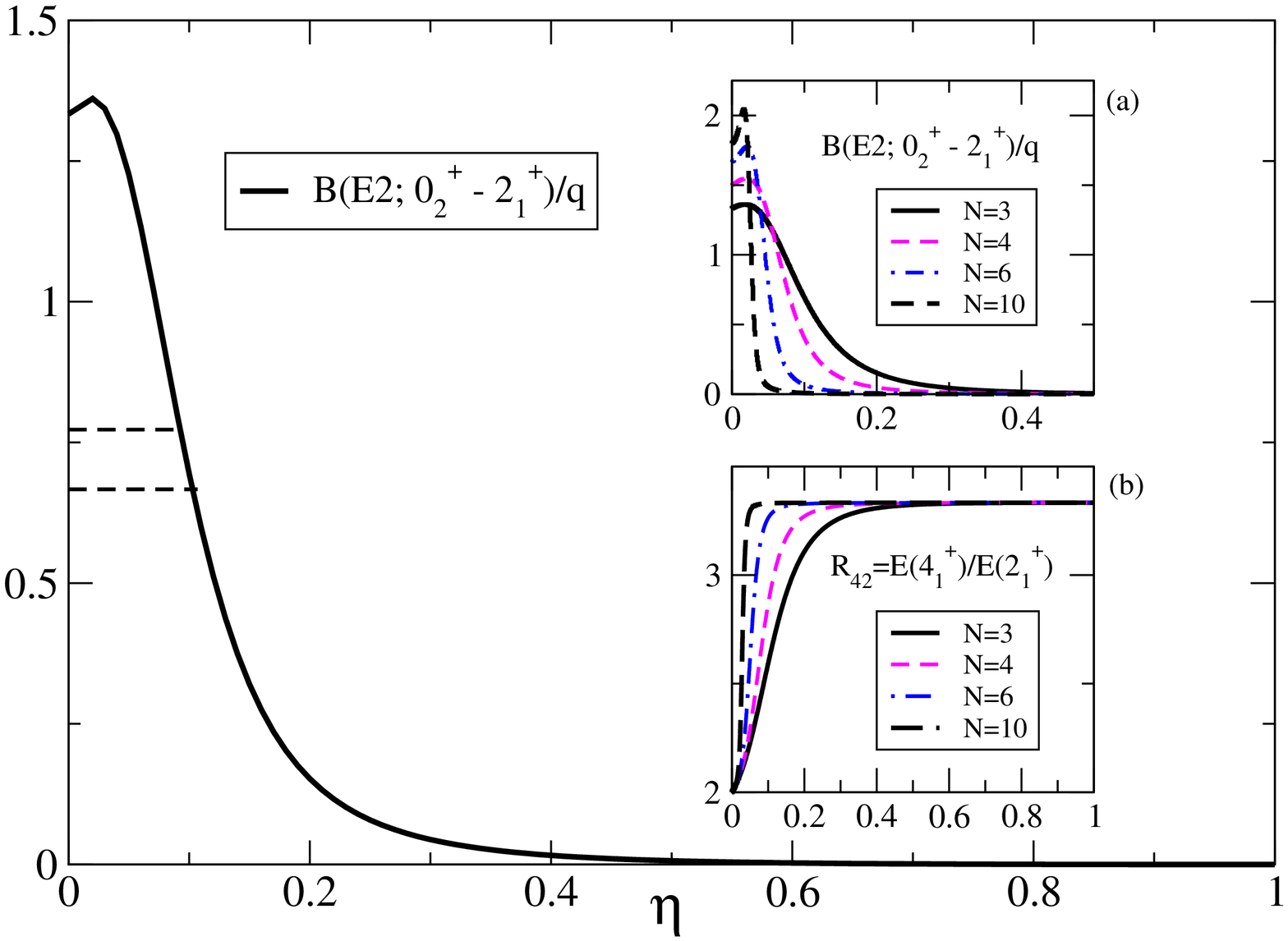}
\bf{Figure 6}
\end{center}
\end{figure}

\section{Conclusions}

In this paper we have proposed a quartet description of $N=Z$ nuclei in a formalism of elementary bosons. As an application we have studied $^{28}$Si.
The microscopic foundation of the quartet bosons has been supported by the outcomes of a mapping procedure.
An analysis of the potential energy surface has evidenced the peculiar nature of $^{28}$Si by placing it, only known case so far, at the critical point of the U(5)-$\overline{\rm SU(3)}$ transition of the IBM structural diagram. Oblate nuclei are rather rare in nature \cite{cejnar} and no sequence of nuclei exhibiting a spherical-oblate transition has ever been observed to our knowledge. This holds true in particular for the small $sd$ shell which hosts $^{28}$Si just in the middle and where, in particular, nuclei differing by one boson quartet from this nucleus exhibit 
a deformed prolate spectrum  on one side ($^{24}$Mg) and a vibrational-like spectrum on the other side ($^{32}$S).
The case of $^{28}$Si thus appears different from those of other well-established critical nuclei such as $^{150}$Nd, $^{152}$Sm 
or $^{154}$Gd which lie along isotopic chains exhibiting a spherical-prolate transition \cite{scholten,iachello,garcia-ramos}. 
Although the present analysis has been limited to $^{28}$Si
we expect that the quartet boson model will represent an useful tool to describe heavier unstable $N=Z$ nuclei, for which new experimental data are 
going to be provided by the radioactive beam facilities.

\section*{Acknowledgements}
The authors wish to thank Prof. F. Iachello for useful discussions.
This work was supported by a grant of Romanian Ministry of Research and Innovation, CNCS - UEFISCDI, project number
PN-III-P4-ID-PCE-2016-0481, within PNCDI III.

\section*{References}

\bibliography{biblio}

\newpage
\section*{Figure captions}

\section*{Figure 1}
\noindent
Experimental \cite{exp} and theoretical low-energy spectra of $^{28}$Si.
Arrows represent B(E2) transitions and the corresponding values (in W.u.) are given by the numbers next to them. The circle on the $2^+$ level stands for the quadrupole moment of this state (in eb). The number below the ground state gives the binding energy (experimental value from Ref.  \cite{brown_tables}).

\section*{Figure 2}
\noindent
(Color online)
Phenemenological and microscopically derived coefficients of the two-body part of the boson Hamiltonian (\ref{1}) (see text).

\section*{Figure 3}
\noindent
(Color online)
Potential energy surface $E(\beta ,\gamma )$ for $^{28}$Si calculated with the Hamiltonian (\ref{1}). The inset shows this energy for $\gamma =60^{\rm o}$ on an expanded scale. 

\section*{Figure 4}
\noindent
(Color online)
Potential energy surface calculated with the Hamiltonian (\ref{11}) for $\eta =\eta_{cr}=0.129$ (in arbitrary units). The inset shows this energy for $\gamma =60^{\rm o}$ on an expanded scale.

\section*{Figure 5}
\noindent
Comparison of spectra generated with the Hamiltonians (\ref{11}) (left) and (\ref{1}) (right). See text for details. Symbols are as in Fig. 1.

\section*{Figure 6}
\noindent
(Color online)
The B(E2; $0^+_2$ - $2^+_1$) transition, normalised to q=B(E2; $2^+_1$ - $0^+_1$), as a
function of $\eta$. The dotted parallel lines intersecting the B(E2) curve delimitate
the range of experimental values. 
Inset (a) shows the B(E2) ratio for various boson numbers N. Inset (b) displays
the ratio $R_{42}=E(4^+_1)/E(2^+_1)$ as a function of $\eta$ for various N. 

\end{document}